\documentclass[seceq]{ptptex}

\preprintnumber[3cm]{
KUNS-2287\\PTPTeX ver.0.8\\ August, 1997}

\title{Master Equations for Gravitational Perturbations\\
 of Static Lovelock Black Holes in Higher Dimensions
}

\author{Tomohiro \textsc{Takahashi} and Jiro \textsc{Soda}
}

\inst{
Department of Physics,  Kyoto University, Kyoto, 606-8502, Japan
}

\abst{
In this paper, we derive the master equations for gravitational perturbations of 
vector and scalar type for static vacuum Lovelock black holes. 
Together with our previous work on the tensor type perturbation, we now provide 
the full set of master equations that governs all types of gravitational perturbations of 
static vacuum Lovelock black holes in any dimensions. 
}

\begin{document}

\maketitle

\section{Introduction}

There is a long history in black hole perturbation theory since the seminal paper 
by Regge and Wheeler.\cite{Regge:1957td,Zerilli:1970se, Chandra} \ 
It is well known that there exist master equations for gravitational
perturbations of static black holes
in 4-dimensions.\cite{Regge:1957td,Zerilli:1970se, Chandra,
Moncrief:1974am,Takahashi:2010th} \ 
Surprisingly, the master equations have been obtained even for stationary
black holes.\cite{Teukolsky:1973ha,Chandra} \ It should be mentioned that
the master equations for gravitational perturbations have played an important
role in gravitational wave physics. 

Recently, higher dimensional black holes have attracted much attention. 
This is because higher dimensional black holes could be created at the 
LHC.\cite{Giddings:2001bu} \ 
In fact, in the context of the braneworld with large extra-dimensions,
the predicted production rate of black holes is within reach of accelerators.
Another reason is that higher dimensional black holes
have been used to analyze strongly coupled finite temperature field theories
through the AdS/CFT correspondence.
Needless to say, the master equations of gravitational perturbations
of black holes in higher dimensions are crucial for the developments
of these subjects.

The master equations for gravitational perturbations
 of higher dimensional static black holes have been 
obtained by Kodama and Ishibashi.\cite{Kodama:2003jz} \ 
For stationary black holes, unfortunately, there exist only partial results.
To investigate the stability of rotating black holes,
a group theoretical method is developed.\cite{Murata:2007gv} \ 
The method is used to obtain the master equations for gravitational perturbations
of squashed black
 holes~\cite{Kimura:2007cr,Ishihara:2008re,Nishikawa:2010zg}
 and 5-dimensional rotating black holes with equal angular momenta.\cite{Murata:2008yx} \ 
The master equations for a special class of 
rotating black holes in more than 5-dimensions are also
studied.\cite{Kunduri:2006qa,Oota:2008uj,Kodama:2008rq} \ 
Still, it is an open issue if the master equations exist for general rotating black
holes and black rings.

We should notice that, in higher dimensions, Einstein theory is not a unique
theory with the second-order differential equations.
Indeed, the most general theory of gravity is
Lovelock theory which is degenerated into Einstein theory in 4-dimensions. 
In fact, Lovelock theory is a natural extension of Einstein theory
in that the Lovelock theory contains terms only up to the second order derivatives
in the equations of motion. 
In Lovelock theory, it is known that
there exist static spherical symmetric black hole solutions~\cite{Wheeler:1985nh} and 
topological black hole solutions.\cite{Cai:2001dz} \  
Hence, it is legitimate to suppose black holes produced at the LHC are 
of this type.\cite{Barrau:2003tk} \ 
Thus, it is important to obtain master equations in order to
study these Lovelock black holes.

The following argument also stresses the importance of Lovelock theory. 
The point is that  black holes are produced 
at the fundamental scale of higher dimensional theories. 
At such high energy, Einstein theory would be no longer valid.
In fact, as is well known, string theory predicts Einstein theory
only in the low energy limit.\cite{Boulware:1985wk} \ In string theory,
  there are higher curvature corrections in addition to
Einstein-Hilbert term.\cite{Boulware:1985wk} \ Thus, it 
is natural to extend gravitational theory into those
 with higher power of curvature in higher dimensions.
  It is Lovelock theory 
that belongs to such class of theories.\cite{Lovelock:1971yv} \

In the case of second order Lovelock theory, the so-called 
Einstein-Gauss-Bonnet theory, the master equation for tensor
perturbations has been obtained.\cite{Dotti:2004sh} \  
The result has been also extended to the scalar and vector 
perturbations.\cite{Gleiser:2005ra} \ 
Although Einstein-Gauss-Bonnet theory is the most general theory
in five and six dimensions, it is not so in more than six dimensions.
For example, when we consider ten dimensional black holes, 
we need to incorporate the fourth order Lovelock term. 
Indeed, when we consider black holes at the LHC, 
it is important to consider these higher order Lovelock terms.\cite{Rychkov:2004sf} \ 
Hence, in this paper, we derive the master equations for gravitational perturbations
 of black holes in any order
Lovelock theory, namely, in any dimensions. 
We have already derived the master equation for tensor perturbations of 
 static Lovelock black holes in any dimensions.\cite{Takahashi:2009dz} \  
In this paper, we derive master equations for vector and scalar perturbations
of static Lovelock black holes.
 
The organization of this paper is as follows.
 In Section \ref{seq:2}, we briefly review Lovelock theory and static Lovelock black hole solutions. 
 In section \ref{seq:3}, we express the linear Lovelock tensor in terms of
 the perturbed Riemann tensor. 
In section \ref{seq:4}, for completeness,
we derive previous results for tensor perturbations.\cite{Takahashi:2009dz}\ 
In section \ref{seq:5}, we derive the master equation for vector perturbations.
In Section \ref{seq:6}, we deduce the master equation for scalar perturbations.
The final section \ref{seq:7} is devoted to the conclusion.

\section{Static Lovelock Black Holes}
\label{seq:2}

In this section, we review Lovelock theory and introduce 
static black hole solutions. 

In Ref.~\citen{Lovelock:1971yv}, the most general symmetric, divergence free rank (1,1) tensor 
is constructed out of a metric and its first and second derivatives.
The corresponding Lagrangian can be constructed from $m$-th order Lovelock terms
\begin{eqnarray}
  {\cal L}_m = \frac{1}{2^m} 
  \delta^{\lambda_1 \sigma_1 \cdots \lambda_m \sigma_m}_{\rho_1 \kappa_1 \cdots \rho_m \kappa_m}
  R_{\lambda_1 \sigma_1}{}^{\rho_1 \kappa_1} \cdots  R_{\lambda_m \sigma_m}{}^{\rho_m \kappa_m}
                       \ ,
\end{eqnarray}
where  $R_{\lambda \sigma}{}^{\rho \kappa}$ is the Riemann tensor in $D$-dimensions
and $\delta^{\lambda_1 \sigma_1 \cdots \lambda_m \sigma_m}_{\rho_1 \kappa_1 \cdots \rho_m \kappa_m}$ is the 
generalized totally antisymmetric Kronecker delta defined by  
\begin{eqnarray}
\delta^{\mu_1\mu_2\cdots \mu_p}_{\nu_1\nu_2\cdots\nu_p}={\rm det}
\left(
\begin{array}{cccc}
\delta^{\mu_1}_{\nu_1}&\delta^{\mu_1}_{\nu_2}&\cdots&\delta^{\mu_1}_{\nu_p}\\
\delta^{\mu_2}_{\nu_1}&\delta^{\mu_2}_{\nu_2}&\cdots&\delta^{\mu_2}_{\nu_p}\\
\vdots&\vdots&\ddots&\vdots\\
\delta^{\mu_p}_{\nu_1}&\delta^{\mu_p}_{\nu_2}&\cdots&\delta^{\mu_p}_{\nu_p}
\end{array}
\right)\ .
\nonumber
\end{eqnarray}
Then, Lovelock Lagrangian in  $D$-dimensions is defined by
\begin{eqnarray}
  L = \sum_{m=0}^{k} c_m {\cal L}_m \ ,   \label{eq:lag}
\end{eqnarray}
where we defined the maximum order $k\equiv [(D-1)/2]$ and  $c_m$ are 
arbitrary constants. 
Here, $[z]$ represents the maximum integer satisfying $[z]\leq z$. 
Hereafter, we set $c_0=-2\Lambda$, $c_1=a_1=1$ and $c_m=a_m/m\ (m\geq 2)$, for convenience. 
Taking variation of the Lagrangian with respect to the metric,
 we can derive the Lovelock equation
\begin{eqnarray}
	0={\cal G}_{\mu}^{\nu}
      =\Lambda \delta_{\mu}^{\nu}-\sum_{m=1}^{k}\frac{1}{2^{(m+1)}}\frac{a_m}{m} 
	 \delta^{\nu \lambda_1 \sigma_1 \cdots \lambda_m \sigma_m}_{\mu \rho_1 \kappa_1 \cdots \rho_m \kappa_m}
       R_{\lambda_1 \sigma_1}{}^{\rho_1 \kappa_1} \cdots  R_{\lambda_m \sigma_m}{}^{\rho_m \kappa_m}  \ . \label{eq:EOM}
\end{eqnarray}

As shown in Refs.~\citen{Wheeler:1985nh} and~\citen{Cai:2001dz}, there exist static 
exact black hole solutions of the Lovelock equations. 
Let us consider the following metric
\begin{eqnarray}
   ds^2=-f(r)dt^2 + \frac{dr^2}{f(r)}+r^2{\bar \gamma}_{i j}dx^idx^j \ ,\label{eq:solution}
\end{eqnarray}
where ${\bar \gamma}_{ij}$ is the metric of $n\equiv D-2$-dimensional constant
 curvature space with a curvature $\kappa$=1,0 or -1.  
Using this metric ansatz, we can calculate Riemann tensor components as
\begin{eqnarray}
	&\ &R_{tr}{}^{tr}=-\frac{f^{''}}{2} \ ,\quad
      R_{ti}{}^{tj}=R_{ri}{}^{rj}=-\frac{f^{'}}{2r}\delta_{i}{}^{j} \ ,\quad \nonumber\\
      &\ &R_{ij}{}^{kl}=\left(\frac{\kappa-f}{r^2}\right)\left(\delta_{i}{}^{k}\delta_{j}{}^{l}-\delta_{i}{}^{l}\delta_{j}{}^{k}\right) \ . \label{eq:riemann}
\end{eqnarray}
Substituting (\ref{eq:riemann}) into (\ref{eq:EOM}) and defining a new variable
$\psi(r)$ by
\begin{eqnarray}
       f(r)=\kappa-r^2\psi(r) \ , \label{eq:def}
\end{eqnarray}
we obtain an algebraic equation 
\begin{eqnarray}
       W[\psi]\equiv\sum_{m=2}^{k}\left[\frac{a_m}{m}\left\{\prod_{p=1}^{2m-2}(n-p)\right\}\psi^m\right]+\psi-\frac{2\Lambda}{n(n+1)}=\frac{\mu}{r^{n+1}}  \ .
\label{eq:poly}
\end{eqnarray}
In (\ref{eq:poly}), we used $n = D-2$ and $\mu$ is a constant of integration
 which is related to the ADM mass of black holes as~\cite{Myers:1988ze}  
\begin{eqnarray}
	M=\frac{2\mu\pi^{(n+1)/2}}{\Gamma((n+1)/2)} \ , \label{eq:ADM}
\end{eqnarray}
where we used a unit $16\pi G=1$.

In the following sections, we study gravitational perturbations around the
general vacuum solutions (\ref{eq:def})
obtained by solving algebraic equation (\ref{eq:poly}).


\section{Linear Lovelock Tensor}
\label{seq:3}

In this section, we present general formulas for the linear Lovelock tensor 
around the solution (\ref{eq:solution}).  
The method introduced in this section is based on symmetry of the static Lovelock black holes. 
In detail, the point is that only $R_{tr}{}^{tr}$, $R_{ti}{}^{tj}$, $R_{ri}{}^{rj}$ and $R_{ij}{}^{kl}$ 
have nonzero value for the Riemann tensor due to the metric ansatz (\ref{eq:solution}).

From now on, we use $\mu,\ \nu,\ \cdots$ for  $(t,r,x^i)$, 
$x,\ y,\ \cdots$ for $(r,x^i)$, and $i,\ j,\  \cdots$ for $(x^i)$, respectively. 
With this notation, we can show that 
\begin{eqnarray}
	\delta_{i_1i_2\cdots i_{m}}^{j_1j_2\cdots j_{m}}\delta_{j_1}^{i_1}=\{n-(m-1)\}\delta_{i_2i_3\cdots i_{m}}^{j_2j_3\cdots j_{m}} 
	\label{formulae}
\label{}
\end{eqnarray}
by induction. This formula is useful for later calculations.
It is also easy to see that the linear Lovelock tensor reads
\begin{eqnarray}
\delta{\cal G}_{\mu}^{\nu}
	 &=&-\sum_{m=1}^{k}\frac{a_m}{2^{(m+1)}} 
	 \delta^{\nu \lambda_1 \sigma_1 \cdots \lambda_m \sigma_m}_{\mu \rho_1 \kappa_1 \cdots \rho_m \kappa_m}\nonumber\\
       &\ &\times R_{\lambda_1 \sigma_1}{}^{\rho_1 \kappa_1} \cdots  R_{\lambda_{m-1} \sigma_{m-1}}{}^{\rho_{m-1} \kappa_{m-1}} \delta R_{\lambda_m \sigma_m}{}^{\rho_m \kappa_m}\label{eq:pert}\ .
\label{}
\end{eqnarray}

Let us explain how to calculate $\delta{\cal G}_t^t$ in detail
and merely present final results for other components.  
First of all, we consider the  totally antisymmetric Kronecker delta 
in $\delta {\cal G}_t^t$, namely,
$\delta^{t \lambda_1 \sigma_1 \cdots \lambda_m \sigma_m}_{t \rho_1 \kappa_1 \cdots \rho_m \kappa_m}$. Because of the antisymmetry, $t$ cannot show up twice in 
the indexes. Hence, this Kronecker delta can be rewritten as 
\begin{eqnarray}
\delta_{t x_1x_2\cdots x_{2m-1}x_{2m}}^{t y_1y_2\cdots y_{2m-1}y_{2m}}= \delta_{x_1x_2\cdots x_{2m-1}x_{2m}}^{y_1y_2\cdots y_{2m-1}y_{2m}}\ ,
\label{}
\end{eqnarray}
where we used properties $\delta_t^t=1$ and $\delta_{x_p}^t=0$. 
Thus, we can rewrite $\delta{\cal G}_t^t$ as 
\begin{eqnarray}
	\delta{\cal G}_t^t&=&\sum_m\left(-\frac{a_m}{2^{m+1}}\right)\delta_{x_1x_2\cdots x_{2m-1}x_{2m}}^{y_1y_2\cdots y_{2m-1}y_{2m}}\nonumber\\
	&\ &\times R_{y_1y_2}{}^{x_1x_2}\cdots  R_{y_{2m-3}y_{2m-2}}{}^{x_{2m-3}x_{2m-2}}\delta R_{y_{2m-1}y_{2m}}{}^{x_{2m-1}x_{2m}}\nonumber \ .
\end{eqnarray}
Here, when taking the summation of $x_p$ and $y_p$, we have to consider three cases.  
One is that there is no $r$ index, that is, all $x_p$ and $y_p$ are $i_p$ and $j_p$, respectively.  
For the other two cases, we have $r$ index. 
Assume $x_p=r$, then, there must exist $p^{'}$ such that $y_{p^{'}}=r$ because 
the totally antisymmetric delta consists of Kronecker delta. 
One possibility is $x_{2q-1}=r$ or $x_{2q}=r$ $(1\leq q\leq m-1)$. 
For this case, $p^{'}$ must be $p^{'}=2q-1\ {\rm or}\ 2q$ 
because of the background property (\ref{eq:riemann}). 
The remaining possibility is $x_{2m-1}=r$ or $x_{2m}=r$. For this case, 
 we have to take $p^{'}=2m-1\ {\rm or}\ 2m$. The reason is as follows. 
If $p^{'}\neq 2m-1$ nor $p^{'}\neq 2m$, there must exist $q^{'}$ $(1\leq q^{'}\leq 2m-2)$ such that $y_{q^{'}}=r$. 
For example, if we take $y_1=r$, $x_1$ or $x_2$ must be $r$ 
because of the formula (\ref{eq:riemann}). 
 In any case, $\delta_{x_1 x_2\cdots}^{y_1y_2\cdots}$ must be zero because 
of the antisymmetry. 
To summarize, $\delta{\cal G}_t^t$ can be written as
\begin{eqnarray}
	\delta{\cal G}_t^t&=&\sum_{m}\left(-\frac{a_m}{2^{m+1}}\right)\Biggl[\nonumber\\
	&\ &\delta_{i_1\cdots i_{2m}}^{j_1\cdots j_{2m}}R_{j_1j_2}{}^{i_1i_2}\times \cdots \times R_{j_{2m-3}j_{2m-2}}{}^{i_{2m-3}i_{2m-2}}\delta R_{j_{2m-1}j_{2m}}{}^{i_{2m-1}i_{2m}}\nonumber\\
	&+&4(m-1)\delta_{i_1\cdots i_{2m-1}}^{j_1\cdots j_{2m-1}}\nonumber\\
	&\ &\times R_{rj_1}{}^{r i_1}R_{j_2j_3}{}^{i_2 i_3}\times \cdots \times R_{j_{2m-4}j_{2m-3}}{}^{i_{2m-4}i_{2m-3}}\delta R_{j_{2m-2}j_{2m-1}}{}^{i_{2m-2}i_{2m-1}}\nonumber\\
	&+&4\delta_{i_1\cdots i_{2m-1}}^{j_1\cdots j_{2m-1}}R_{j_1j_2}{}^{i_1i_2}\times\cdots\times R_{j_{2m-3}j_{2m-2}}{}^{i_{2m-3}i_{2m-2}}\delta R_{j_{2m-1}r}{}^{i_{2m-1}r}\Biggr]\ . \label{tt_decompose}
\end{eqnarray}   
Substituting  the background quantities (\ref{eq:riemann}) into (\ref{tt_decompose}) and using the formula (\ref{formulae}), we can proceed as 
\begin{eqnarray}
	\delta{\cal G}_t^t&=&\sum_{m=1}^{k}\left(-\frac{a_m}{2^{m+1}}\right)\Biggl[\nonumber\\
	&\ &2^{m-1}\left(\frac{\kappa-f}{r^2}\right)^{m-1}\delta_{i_1\cdots i_{2m}}^{j_1\cdots j_{2m}}\delta_{j_1}^{i_1}\cdots \delta_{j_{2m-2}}^{i_{2m-2}}\delta R_{j_{2m-1}j_{2m}}{}^{i_{2m-1}i_{2m}}\nonumber\\
	&\ &+4(m-1)2^{m-2}\left(-\frac{f^{'}}{2r}\right)\left(\frac{\kappa-f}{r^2}\right)^{m-2}\nonumber\\
	&\ &\hspace{1cm}\times\delta_{i_1\cdots i_{2m-1}}^{j_1\cdots j_{2m-1}} \delta_{j_1}^{i_1}\cdots \delta_{j_{2m-3}}^{i_{2m-3}}\delta R_{j_{2m-2}j_{2m-1}}{}^{i_{2m-2}i_{2m-1}}\nonumber\\
	&\ &+4\cdot2^{m-1}\left(\frac{\kappa-f}{r^2}\right)^{m-1}\delta_{i_1\cdots i_{2m-1}}^{j_1\cdots j_{2m-1}}\delta_{j_1}^{i_1}\cdots \delta_{j_{2m-2}}^{i_{2m-2}}\delta R_{j_{2m-1}r}{}^{i_{2m-1}r}\Biggr]\nonumber
	\end{eqnarray}
	\begin{eqnarray}
	&=&\sum_{m=1}^{k} a_m\Biggl[\Biggl(-\frac{1}{4}\biggl\{\prod_{p=2}^{2m-1}(n-p)\biggr\}\left(\frac{\kappa-f}{r^2}\right)^{m-1}\nonumber\\
	&\ &\hspace{2cm}+(m-1)\biggl\{\prod_{p=2}^{2m-2}(n-p)\biggr\}\left(\frac{f^{'}}{4r}\right)\left(\frac{\kappa-f}{r^2}\right)^{m-2}\Biggr)
	\delta_{ik}^{jl} \delta R_{jl}{}^{ik}\nonumber\\
	&\ &\hspace{2.5cm}-\biggl\{\prod_{p=1}^{2m-2}(n-p)\biggr\}\left(\frac{\kappa-f}{r^2}\right)^{m-1}\delta_i^j\delta R_{jr}{}^{ir}\Biggr]\nonumber\\
	&=&-\frac{T^{'}}{2(n-1)r^{n-2}}\delta_i^j\delta_k^l\delta R_{jl}{}^{ik}-\frac{T}{r^{n-1}}\delta_i^j\delta R_{jr}{}^{ir}\ ,  \label{}
\label{}
\end{eqnarray}
 where we used the relation $\delta_{ij}^{kl}\delta R_{kl}{}^{ij}
 =2\delta_{i}^{k}\delta_{j}^{l}\delta R_{kl}{}^{ij}$ in the last equality. 
 Here, $T(r)$ is defined by 
\begin{eqnarray}
	T(r)\equiv r^{n-1}\partial_{\psi}W[\psi]=r^{n-1}\left(1+\sum_{m=2}^{k}\left[a_m\left\{\prod_{p=1}^{2m-2}(n-p)\right\}\psi^{m-1}\right]\right) \ .
	\label{def_of_T}
\end{eqnarray}

Similarly, we can deduce other components of the linear Lovelock tensor. 
The results are as follows;
\begin{eqnarray}
	\delta{\cal G}_t^t&=&-\frac{T^{'}}{2(n-1)r^{n-2}}\delta_i^j\delta_k^l\delta R_{jl}{}^{ik}-\frac{T}{r^{n-1}}\delta_i^j\delta R_{jr}{}^{ir}\ ,\nonumber\\
	\delta{\cal G}_t^r&=&-\frac{T}{r^{n-1}}\delta_i^j\delta R_{jt}{}^{ir}\ ,\nonumber\\
	\delta{\cal G}_t^i&=&\frac{T^{'}}{(n-1)r^{n-2}}\delta_k^j\delta R_{tj}{}^{ik}+\frac{T}{r^{n-1}}\delta R_{tr}{}^{ir}\ ,\nonumber\\
	\delta{\cal G}_r^r&=&-\frac{T^{'}}{2(n-1)r^{n-2}}\delta_i^j\delta_k^l\delta R_{jl}{}^{ik}-\frac{T}{r^{n-1}}\delta_i^j\delta R_{jt}{}^{it}\ ,\nonumber\\
	\delta{\cal G}_r^i&=&\frac{T^{'}}{(n-1)r^{n-2}}\delta_k^j\delta R_{rj}{}^{ik}+\frac{T}{r^{n-1}}\delta R_{rt}{}^{it}\ ,\nonumber\\
	\delta{\cal G}_i^j&=&\frac{T^{'}}{(n-1)r^{n-2}}\left(\delta R_{ti}{}^{tj}+\delta R_{ri}{}^{rj}\right)+\frac{T^{''}}{(n-1)(n-2)r^{n-3}}\delta_l^k \delta R_{ik}{}^{jl}\nonumber\\
	                  &\ &-\delta_i^j\Biggl[\frac{T}{r^{n-1}}\delta R_{tr}{}^{tr}+\frac{T^{'}}{(n-1)r^{n-2}}\left(\delta R_{tk}{}^{tl}+\delta R_{rk}{}^{rl}\right)\delta_l^k\nonumber\\
	                  &\ &\hspace{2.5cm}+\frac{T^{''}}{2(n-1)(n-2)r^{n-3}}\delta_l^k\delta_p^q \delta R_{kq}{}^{lp}
	\Biggr]\ .
	\label{delta_G}
\end{eqnarray} 
Thus, in order to derive the linear Lovelock equations, we need to know the 
perturbed Riemann tensor $\delta R_{\mu\nu}{}^{\rho \lambda}$. 

Since Lovelock black holes have $n$-dimensional symmetric space, we can classify
 perturbations into tensor, vector and scalar type perturbations. 
In following sections, we treat these three type of perturbations separately.

\section{Master equation for tensor perturbations}
\label{seq:4}

In this section, we derive the master equation for tensor perturbations
of static Lovelock black holes. 

Tensor perturbations around the solution (\ref{eq:solution}) is characterized by
\begin{eqnarray}
	\delta g_{a b}=0 \ , \quad \delta g_{a i}=0 \ ,\quad 
      \delta g_{i j}=r^2 \phi(t,r)\bar{h}_{i j}(x^i ) \ , 
      \label{metric:tensor}
\end{eqnarray}
where $a,b=(t,r)$ and $\phi (t,r)$ represents the dynamical degrees of freedom.
Here, tensor harmonics $\bar{h}_{ij}$  are defined by
\begin{eqnarray}
	\bar{\nabla}^{k}\bar{\nabla}_{k} \bar{h}_{ij}=-\gamma_t \bar{h}_{ij} \ , \qquad
	\bar{\nabla}^{i} \bar{h}_{ij}=0 \ ,\quad \bar{\gamma}^{ij}\bar{h}_{ij}=0\ , 
\end{eqnarray}
where  $\bar{\nabla}^{i}$ denotes a covariant derivative with respect to
 $\bar{\gamma}_{ij}$ and the eigenvalue is given by 
$\gamma_t =\ell (\ell +n-1)-2$, ($\ell =2,3,4 \cdots$) for $\kappa=1$ 
and  positive real numbers for $\kappa=-1,0$. Note that indexes $i,\ j,\cdots$ are raised or lowered by ${{\bar\gamma}_{ij}}$. 

For tensor perturbations, from (\ref{delta_G}), it is clear that components other 
than $\delta {\cal G}_i^j$ vanish and 
the terms proportional to $\delta_i^j$ in $\delta {\cal G}_i^j$ also disappear. Therefore, what we have to calculate are 
$\delta R_{ti}{}^{tj}$, $\delta R_{ri}{}^{rj}$ and $ \delta^m_l \delta R_{im}{}^{jl}$. 
From the metric ansatz (\ref{eq:solution}) and (\ref{metric:tensor}), 
these components can be deduced as
\begin{eqnarray}
	\delta R_{ti}{}^{tj}
      &=&\left[\frac{{\ddot \phi}}{2f}-\frac{f^{'}\phi^{'}}{4}\right] {\bar h}_{i}{}^{j}
      \ , \nonumber\\ 
      \delta R_{ri}{}^{rj}&=&\left[-\frac{f\phi^{''}}{2}+\left(-\frac{f^{'}}{4}-\frac{f}{r}\right)\phi^{'}\right] {\bar h}_{i}{}^{j} \ , \nonumber\\
      \delta^m_l \delta R_{im}{}^{jl}&=& 
      \left[ -\frac{n-2}{2} \frac{f}{r} \phi' 
           +\frac{2\kappa +\gamma_t}{2r^2} \phi \right]{\bar h}_{i}{}^{j}
      \ .\label{eq:deltariemann}
\end{eqnarray}
Then, substituting these results into (\ref{delta_G}), 
we can calculate $\delta{\cal G}_i^j$ as follows; 
\begin{eqnarray}
	&\ &(n-1)r^{n-2}\delta{\cal G}_{i}{}^{j}\nonumber\\
	&\ &\hspace{0.3cm}=T^{'}\left(\delta R_{ti}{}^{tj}+\delta R_{ri}{}^{rj}\right)+\frac{rT^{''}}{(n-2)}\delta^m_l \delta R_{im}{}^{jl}\nonumber\\
	                     &\ & \hspace{0.3cm}=\left[\frac{T^{'}}{2f}\left({\ddot \phi}-f^2\phi^{''}\right)-\left(\frac{f^{'}T^{'}}{2}+\frac{fT^{'}}{r}+\frac{fT^{''}}{2}\right)\phi^{'}+\frac{(2\kappa+\gamma_t)T^{''}}{2(n-2)r}\phi\right]{\bar h}_{i}{}^j\ .\label{eq:deltaG}
\end{eqnarray}
 
Separating the variables $\phi(r,t)=\chi(r)e^{-i\omega t}$, 
we can derive the master equation for the tensor perturbations 
from the linear Lovelock equation $\delta{\cal G}_{\mu}^{\nu}=0$ as follows; 
\begin{eqnarray}
	- f^2\chi^{''}
      - \left( f^2 \frac{T^{''}}{T^{'}}+ \frac{2f^2}{r}
      +  f f^{'} \right) \chi^{'}
      +  \frac{(2\kappa+\gamma_t)f}{(n-2)r}\frac{T^{''}}{T^{'}} \chi =   \omega^2 \chi \ .
 \label{eq:master_eq}
\end{eqnarray}
Here, we should stress that
we have assumed nothing for Lovelock coefficients and $f(r)$. 
Hence, the master equation we derived is quite general. 

Furthermore,  introducing a new function 
$\Psi(r)=\chi(r)r\sqrt{T^{'}(r)} $ and using tortoise coordinate $r^{*}$ which is defined as $dr^{*}/dr=1/f(r)$, 
we can transform the master equation (\ref{eq:master_eq}) 
into Schr${\ddot{\rm o}}$dinger type equation  
\begin{eqnarray}
	-\frac{d^2\Psi}{dr^{*2}}+V_t(r(r^*))\Psi=\omega^2\Psi  \ , 
\end{eqnarray}
where we have defined the effective potential
\begin{eqnarray}
	V_t(r)=\frac{(2\kappa+\gamma_t)f}{(n-2)r}\frac{d \ln{T^{'}}}{dr}+\frac{1}{r\sqrt{T^{'}}}f\frac{d}{dr}\left(f\frac{d}{dr}r\sqrt{T^{'}}\right) \label{eq:potential} \ .
\end{eqnarray}
Here, we have assumed $T^{'}>0$ in order to avoid ghost instability.\cite{Takahashi:2009dz} \ 

\section{Master equation for vector perturbations}
\label{seq:5}

In this section, we consider vector perturbations of static Lovelock black holes 
and derive the master equation. 

We take the Regge-Wheeler gauge 
\begin{eqnarray}
	\delta g_{\mu \nu}=
	\left(
	\begin{array}{cc|c}
	0&0&v_i\\
	0&0&w_i\\ \hline
	{\bf sym}&{\bf sym}&{\bf 0}
	\end{array}
	\right)\label{vector_RW}  \ ,
\end{eqnarray}
where $v_i$ and $w_i$ satisfy transverse condition ${\bar \nabla}^{i}v_i={\bar \nabla}^iw_i=0$ and ``sym" means symmetric part of metric perturbations.  

From the ansatz (\ref{vector_RW}), 
it is clear that $\delta {\cal G}_t^t=\delta{\cal G}_r^r=\delta{\cal G}_t^r=0$. 
The components of $\delta R_{\mu\nu}{}^{\rho\lambda} $
 which is necessary for calculations of non-zero components of $\delta {\cal G}_\mu^\nu$ 
 are as follows;
\begin{eqnarray}
	\delta R_{tk}{}^{ij}&=&\frac{\kappa-f}{r^4}\left(\delta_k^j v^i-\delta_k^iv^j\right)+\frac{f^{'}}{2r^3}\left(\delta_k^j v^i-\delta_k^iv^j\right)\nonumber\\
	                    &\ &-\frac{1}{2r^4}\left[-v^j{}_{|k}{}^{|i}+v^i{}_{|k}{}^{|j}+\kappa(\delta_k^j v^i-\delta_k^iv^j)\right]\nonumber\\
	                    &\ &-\frac{f}{2r^3}\Biggl[\delta_k^j\left(v^{i'}-{\dot w}^i-\frac{2}{r}v^{i}\right)
	                               -\delta_k^i\left(v^{j'}-{\dot w}^j-\frac{2}{r}v^{j}\right)\Biggr]\ ,\nonumber\\                      
	\delta R_{tr}{}^{ir}&=&-\frac{f^{'}}{2r^3}v^i+\frac{f^{''}}{2r^2}v^{i}\nonumber\\
	&\ &-\frac{f}{2r^3}\left(v^{i'}-{\dot w}^i-\frac{2}{r}v^{i}\right)-\frac{f}{2r^2}\left(v^{i'}-{\dot w}^i-\frac{2}{r}v^{i}\right)^{'}\ ,\nonumber\\	                    
	\delta R_{rk}{}^{ij}&=&\frac{\kappa-f}{r^4}\left(\delta_k^j w^i-\delta_k^iw^j\right)+\frac{f^{'}}{2r^3}\left(\delta_k^j w^i-\delta_k^iw^j\right)\nonumber\\
	                    &\ &\hspace{1cm}+\frac{1}{2r^4}\left[w^j{}_{|k}{}^{|i}-w^i{}_{|k}{}^{|j}-\kappa(\delta_k^j w^i-\delta_k^iw^j)\right]\ ,\nonumber\\
	\delta R_{tr}{}^{ti}&=&-\frac{f^{'}}{2r^3}w^i+\frac{f^{''}}{2r^2}w^{i}-\frac{1}{2r^2f}\left(v^{i'}-{\dot w}^i-\frac{2}{r}v^{i}\right)^{\cdot}\ , \nonumber\\
	\delta R_{ti}{}^{tj}&=&-\frac{1}{2r^2f}\left({\dot v}_i{}^{|j}+{\dot v}^j{}_{|i}\right)+\frac{f^{'}}{4r^2}\left(w_i{}^{|j}+w^j{}_{|i}\right)\ ,\nonumber
	\end{eqnarray}
	                               \begin{eqnarray}
	\delta R_{ri}{}^{rj}&=&\frac{f^{'}}{4r^2}\left(w_i{}^{|j}+w^j{}_{|i}\right)+\frac{f}{2r^2}\left(w_i{}^{|j}+w^j{}_{|i}\right)^{'}\ ,\nonumber\\
	\delta R_{ij}{}^{kl}&=&\frac{f}{2r^3}\Biggl[
	\delta_j^l\left(w_i{}^{|k}+w^k{}_{|i}\right)-\delta_j^k\left(w_i{}^{|l}+w^l{}_{|i}\right)\nonumber\\
	&\ &\hspace{1cm}+\delta_i^k\left(w_j{}^{|l}+w^l{}_{|j}\right)-\delta_i^l\left(w_j{}^{|k}+w^k{}_{|j}\right)
	\Biggr]\ ,
\end{eqnarray} 
where we use $|i$ as a covariant derivative with respect to ${\bar \gamma}_{ij}$ instead of ${\bar \nabla}_i$. 
Then, substituting these results into (\ref{delta_G}), we can get the linear
 Lovelock tensor as follows:
\begin{eqnarray}
	2r^{n+2}\delta {\cal G}_t^i&=&-\frac{T^{'}}{n-1}v^{i}{}^{|k}{}_{|k}-\kappa T^{'}v^{i}-f\left\{rT\left(v^{i'}-{\dot w}^{i}-\frac{2}{r}v^{i}\right)\right\}^{'}\ ,\nonumber\\
	2r^{n+2}\delta {\cal G}_r^i&=&-\frac{T^{'}}{n-1}w^{i}{}^{|k}{}_{|k}-\kappa T^{'}w^{i}-\frac{rT}{f}\left(v^{i'}-{\dot w}^{i}-\frac{2}{r}v^{i}\right)^{\cdot}\ , \nonumber\\
	2(n-1)r^{n}\delta{\cal G}_i^j&=&-\frac{T^{'}}{f}\left({\dot v}_i{}^{|j}+{\dot v}^j{}_{|i}\right)+\left\{fT^{'}\left(w_i{}^{|j}+w^j{}_{|i}\right)\right\}^{'}\ ,
	\label{}
\end{eqnarray} 
where we used the relation 
\begin{eqnarray}
\{rf^{'}+2(\kappa-f)\}T=(n+1)\mu\ ,
\label{T_formula}
\end{eqnarray}
and its derivative with respect to $r$. 
Then, after expanding metric perturbations by vector harmonics ${\bar V}_i$, which satisfies ${\bar \nabla}_k{\bar \nabla}^k {\bar V}_i=-\gamma_v{\bar V}_i$ with
$\gamma_v=\ell(\ell+n-1)-1$ $(\ell\geq 2)$ for $\kappa=1$ and 
 non-negative numbers for $\kappa=0,-1$ , we can 
get the linear Lovelock equation as 
\begin{eqnarray}
	\left\{
	\begin{array}{l}
	\left(\frac{\gamma_v}{n-1}-\kappa \right)vT^{'}-f\left\{rT\left(v^{'}-{\dot w}-\frac{2}{r}v\right)\right\}^{'}=0\ ,\\
	\left(\frac{\gamma_v}{n-1}-\kappa \right)wT^{'}-\frac{rT}{f}\left(v^{'}-{\dot w}-\frac{2}{r}v\right)^{\cdot}=0\ ,\\
	-\frac{T^{'}}{f}{\dot v}+\left(fT^{'}w\right)^{'}=0\ .
	\end{array}
	\right. \label{3}
\end{eqnarray}
Note that only two equations among these equations are independent. 
In fact, we can get the third equation in (\ref{3}) by 
combining a derivative of the first equation with respect to $t$ and
 a derivative of the second equation multiplied by $f(r)$ with respect to $r$. 
Here, we use the second and third equations in (\ref{3}).  
We have not considered the exceptional mode $\gamma_v=(n-1)\kappa$ since its treatment is well known . ~\cite{Kodama:2003jz,Kodama:2000fa}

In this paper, we assume $T^{'}$ is always positive otherwise
 there exists ghost instability for tensor perturbations. 
Under this assumption, we can eliminate $v$ from the second equation in (\ref{3}) 
using the third equation. The resultant master equation is given by 
\begin{eqnarray}
\frac{rT}{f}{\ddot w}-\frac{r^3T}{f}\left\{\frac{f}{r^2T^{'}}\left(fT^{'}w\right)^{'}\right\}^{'}+\left(\frac{\gamma_v}{n-1}-\kappa\right)T^{'}w=0\ .
\label{totyu}
\end{eqnarray}

Furthermore, under the assumption of positivity of $T^{'}$,  we can introduce a 
new variable 
\begin{eqnarray}
	\chi=\frac{f}{r}\sqrt{T^{'}}w\ . \nonumber
\end{eqnarray}
We also perform Fourier transformation $\chi=\Psi e^{-i\omega t}$.
Thus, using the new variable and a tortoise coordinate $r^{*}$, 
 we can rewrite Eq. (\ref{totyu}) as  
\begin{eqnarray}
	-\partial_{r^{*}}^2\Psi+V_v(r)\Psi=\omega^2\Psi  \label{}
\end{eqnarray}     
where we have defined the effective potential for vector perturbations
\begin{eqnarray}
V_v(r)=r\sqrt{T^{'}}f\partial_r \left(f\partial_r\frac{1}{r\sqrt{T^{'}}}\right)+\left(\frac{\gamma_v}{n-1}-\kappa\right)\frac{fT^{'}}{rT}\ .
\label{}
\end{eqnarray}
Note that we have not assumed anything for $f(r)$ and Lovelock coefficients $a_m$
except for the positivity of $T^{'}$. 
In this sense, this Schr${\rm {\ddot o}}$dinger type master equation is quite general.

\section{Master equation for Scalar Perturbations}
\label{seq:6}

In this section, we derive the master equation for scalar perturbations
of static Lovelock black holes. 

We use the Zerilli gauge:
\begin{eqnarray}
	h_{\mu \nu}=
	\left(
	\begin{array}{cc|c}
	f{\bar H}&H_1&0\\
	sym&H/f&0\\ \hline
	{\bf sym}&{\bf sym}&r^2K{\bar \gamma}_{ij}
	\end{array}
	\right) \ ,
\end{eqnarray}
where ``sym" represents symmetric part of metric perturbations.

Following Ref.~\citen{Gleiser:2005ra}, we derive the master equation 
from $\delta {\cal G}_i^j=0$$(i\neq j)$,  $\delta {\cal G}_t^r=0$, $\delta {\cal G}_t^t=0$,
 $\delta {\cal G}_r^i=0$ and $\delta {\cal G}_r^r=0$. 
 Taking look at Eqs.(\ref{delta_G}),
we see that we need $\delta R_{ti}{}^{tj}$, $\delta R_{ri}{}^{rj}$, 
$\delta R_{ij}{}^{kl}$, $\delta R_{ti}{}^{rj}$, $\delta R_{ri}{}^{jk}$ and
 $\delta R_{tr}{}^{ti}$  . 
With the  Zerilli gauge, these components  can be calculated as follows:
\begin{eqnarray}
	\delta R_{ti}{}^{tj}&=&\frac{1}{2r^2}{\bar H}_{|i}{}^{|j}+\left(-\frac{1}{r}{\dot H}_1+\frac{1}{2f}{\ddot K}-\frac{f^{'}}{4}K^{'}
	+\frac{f}{2r}{\bar H}^{'}+\frac{f^{'}}{2r}H\right)\delta_{i}^{j}\ ,\nonumber\\
      \delta R_{ri}{}^{rj}&=&-\frac{1}{2r^2} H_{|i}{}^{|j}+\left(\frac{1}{2r}(fH)^{'}-\frac{f^{'}}{4}K^{'}
	-\frac{f}{r}K^{'}-\frac{f}{2}K^{''}\right)\delta_{i}^{j}\ ,\nonumber\\
	\delta R_{ij}{}^{kl}&=&\frac{1}{2r^2}\left[K_{|j}{}^{|k}\delta_{i}^{l}-K_{|i}{}^{|k}\delta_{j}^{l}-K_{|j}{}^{|l}\delta_{i}^{k}+K_{|i}{}^{|l}\delta_{j}^{k}\right]\nonumber\\
	                    &\ &\hspace{1cm}+\left(-\frac{\kappa}{r^2}K+\frac{f}{r^2}H-\frac{f}{r}K^{'}\right)\left(\delta_i^k\delta_j^l-\delta_i^l\delta_j^k\right)\ ,\nonumber
	                    \end{eqnarray}
	                               \begin{eqnarray}
	\delta R_{ti}{}^{rj}&=&-\frac{f}{2r^2}H_{1|i}{}^{|j}-\left(-\frac{f}{2r}{\dot H}-\frac{f^{'}}{4}{\dot K}+\frac{f}{2}{\dot K}^{'}+\frac{f}{2r}{\dot K}\right)\delta_i^j\ ,\nonumber\\
	\delta R_{ri}{}^{jk}&=&\frac{1}{2r^3}\left(H^{|j}\delta_i^k-H^{|k}\delta_i^j\right)-\frac{1}{2r^2}\left(K^{'|j}\delta_i^k-K^{'|k}\delta_i^j\right)\nonumber\ ,\\
	\delta R_{tr}{}^{ti}&=&\frac{1}{2r^2}\left[\frac{f^{'}}{2f}\left(H+{\bar H}\right)+{\bar H}^{'}-\frac{1}{r}{\bar H}-\frac{1}{f}{\dot H}_1\right]^{|i}\ ,
	\label{Riemann}
\end{eqnarray}
where $|i$ means a covariant derivative with respect to ${\bar \gamma}_{ij}$. 
Then, we can calculate the linear Lovelock equations. 
Expanding metric perturbations in terms of scalar harmonics ${\bar Y}$ 
which satisfy ${\bar \nabla}_k{\bar \nabla}^k{\bar Y}=-\gamma_s {\bar Y}$ with $\gamma_s=\ell(\ell+n-1)$ for 
$\kappa=1$ and positive numbers for $\kappa=0,-1$, 
we obtain the linear Lovelock equations  
\begin{eqnarray}
	 \delta {\cal G}_i^j=0\ (i\neq j)\Leftrightarrow T^{'}({\bar H}-H)=rT^{''}K\ ,\label{ij}
\end{eqnarray}
\begin{eqnarray}
	\delta {\cal G}_t^r=0 \Leftrightarrow -\gamma_s H_1 +n\left[r(K-H)+r^2K^{'}-\frac{r^2f^{'}}{2f}K\right]^{\cdot}=0\ , \label{tr}
\end{eqnarray}
\begin{eqnarray}
	\delta {\cal G}_t^t=0&\Leftrightarrow& \left\{-\gamma_s T-nr(fT)^{'}\right\}H-nTrfH^{'}+rT^{'}(n\kappa-\gamma_s)K\nonumber\\
	                     &\ &\hspace{0.5cm}+\left\{\frac{nTr^2f^{'}}{2}+nf(r^2T)^{'}\right\}K^{'}+nTr^2fK^{''}=0\ ,\label{tt}
\end{eqnarray}
\begin{eqnarray}
	\delta {\cal G}_r^i=0&\Leftrightarrow& rT^{'}\left(\frac{1}{r}H-K^{'}\right)\nonumber\\ &\ &\hspace{0.5cm}+T\left[\frac{f^{'}}{2f}(H+{\bar H})+{\bar H}^{'}-\frac{1}{r}{\bar H}-\frac{1}{f}{\dot H}_1\right]=0\ , \label{ri}
\end{eqnarray}
\begin{eqnarray}
	\delta {\cal G}_r^r=0&\Leftrightarrow& \frac{nTr^2}{f}\left(\frac{2f}{r}{\dot H}_1-{\ddot K}\right)+rT^{'}(n\kappa-\gamma_s)K+\left(nT^{'}r^2f+\frac{nTr^2f^{'}}{2}\right)K^{'}\nonumber\\
	&\ &\hspace{0.5cm}-nr(Tf)^{'}H+\gamma_sT{\bar H}-nTrf{\bar H}^{'}=0\ . \label{rr}
\end{eqnarray}

Now we derive the master equation for scalar perturbations from these equations. 
We do not consider exceptional gauge dependent modes $\gamma_s = 0$ and $\gamma_s = n\kappa$ modes since treatment of these modes is well known.~\cite{Kodama:2003jz,Kodama:2000fa}
We also assume $T^{'}>0$ outside the horizon again. 
First of all, we show that all perturbative variables ${\bar H}$, $H_1$, $H$ and $K$ can be expressed by a single master function $\phi$ defined later. 
From Eq. (\ref{ij}), ${\bar H}$ can be expressed by $H$ and $K$ as 
\begin{eqnarray}
	{\bar H}=H+\frac{rT^{''}}{T^{'}}K 
      \label{barH}   \ .
\end{eqnarray}
By inspecting Eq. (\ref{tr}), we see that it is convenient to
 define a master function $\phi$ as 
\begin{eqnarray}
	H_1\equiv \frac{r}{f}\left({\dot \phi}+{\dot K}\right)\ . \label{H1}
\end{eqnarray}
Then, we can express $H$ using $\phi$ and $K$ by integrating (\ref{tr}) with respect to $t$. The result is given by
\begin{eqnarray}
	H=-\frac{\gamma_s}{nf}\phi+rK^{'}-\frac{A(r)}{2nf}K \label{H}\ ,
\end{eqnarray}
where
\begin{eqnarray}
	A(r)=-2nf+2\gamma_s+nrf^{'}\ . \label{}
\end{eqnarray} 
Note that there may exist an arbitrary function of $r$ as a constant of integration.
 However, this function can be absorbed into the definition of $\phi$. 
From Eqs.(\ref{barH}), (\ref{H1}) and (\ref{H}), it turns out that 
we need to express $K$ by $\phi$ 
in order to express all variables in terms of $\phi$. 
Substituting (\ref{H}) into (\ref{tt}), we obtain such a formula
\begin{eqnarray}
	K=-\frac{2}{A}\left[nrf\phi^{'}+\left(\gamma_s+nrf\frac{T^{'}}{T}\right)\phi\right]
      \ ,    \label{K}
\end{eqnarray} 
where we used a relation 
\begin{eqnarray}
	(AT)^{'}=2\gamma_s T^{'}+n\{(rf^{'}-2f)T\}^{'}=2(\gamma_s-n\kappa)T^{'} \ ,\label{}
\end{eqnarray}
which can be derived from (\ref{T_formula}). 

Now, we are in a position to derive the master equation for the master variable $\phi$.
From Eqs.(\ref{ri}) and (\ref{rr}), we can make the following combination 
\begin{eqnarray}
	&\ &nfr\times({\rm l.h.s\ of\ }(\ref{ri}))+({\rm l.h.s\ of\ }(\ref{rr}))\nonumber\\
	&=&\frac{AT}{2}{\bar H}-(\gamma_s-n\kappa)rT^{'}K+\frac{nTrf^{'}}{2}\left(rK^{'}-H\right)+\frac{nTr^2}{f}\left(\frac{f}{r}{\dot H}_1-{\ddot K}\right) \ . \label{pre-master}
\end{eqnarray} 
Substituting (\ref{barH}), (\ref{H1}), (\ref{H}) and (\ref{K}) into the
above equation (\ref{pre-master}), we get the master equation for scalar perturbations 
\begin{eqnarray}
{\ddot \phi}-f^2\phi^{''}+f^2\left(\ln\left(\frac{A^2}{r^2fT^{'}}\right)\right)^{'}\phi^{'}+Q\phi=0\ ,
\label{scalar_phi}
\end{eqnarray}
where we have defined 
\begin{eqnarray}
Q=\frac{f}{nTr^2}\left[\left(2\frac{(AT)^{'}}{AT}-\frac{T^{''}}{T^{'}}\right)(\gamma_srT+nr^2fT^{'})-n(r^2fT^{'})^{'}\right]\ .
\label{}
\end{eqnarray}
Finally, we change the normalization of $\phi$ as
\begin{eqnarray}
\phi=N(r)\chi  \ , \quad  N=\frac{A}{r\sqrt{T^{'}}} \ .
\label{}
\end{eqnarray}
It is also convenient to move on to
 Fourier space as $\chi=\Psi e^{-i\omega t}$.
Substituting these into the master equation (\ref{scalar_phi}) and
using a tortoise coordinate $r^{*}$, 
we obtain Schr${\ddot{\rm o}}$dinger type equation
\begin{eqnarray}
-\partial_{r^*}^2\Psi+V_s(r)\Psi=\omega^2\Psi    \ ,
\label{}
\end{eqnarray}
where we have defined the effective potential for scalar perturbations
\begin{eqnarray}
V_s(r)&=&2\gamma_sf\frac{(rNT)^{'}}{nNTr^2}\nonumber\\
&\ &-\frac{f}{N}\partial_r\left(f\partial_rN \right)
+2f^2\frac{N^{'2}}{N^{2}}-\frac{f}{T}\partial_r(f\partial_rT)
+2f^2\frac{T^{'2}}{T^2}+2f^2\frac{N^{'}T^{'}}{NT}  \ .
\label{}
\end{eqnarray}
We assumed that $T^{'}$ is always positive because tensor perturbation has 
ghost instability if this assumption is not fulfilled. However, except for this assumption, 
we have not imposed any conditions on the Lovelock coefficient $a_m$ and $f(r)$. 
Therefore, the master equation we have derived is again quite general.

\section{Conclusion}
\label{seq:7}

We have succeeded in deriving master equations for gravitational perturbations
of static Lovelock black holes. The results can be regarded as a generalization
of master equations in Einstein theory derived by Kodama and Ishibashi.
Of course, it is possible to extend our analysis to charged black holes
and other black holes.\cite{Maeda:2010bu,Dotti:2010bw} \ 
Moreover, our result would serve a starting point for studying stationary black holes
 in Lovelock theory. It is also interesting to consider
 Euclidean version of our results in the light of
  black hole thermodynamics.\cite{Kastor:2010gq} \ 
  
 There are many applications of our master equations.
 In an accompanying paper, using the master equations, 
 we show that static Lovelock black holes with small masses are unstable
 in the asymptotically flat cases.\cite{TS} \  
 The application to asymptotically AdS cases is also interesting
 from the point of view of the AdS/CFT 
 correspondence,\cite{Ge:2009ac,Shu:2009ax,deBoer:2009gx} \ 
 in particular, in relation to stability of holographic 
 superconductors.\cite{Gregory:2009fj,Kanno:2010pq} \ 

\section*{Acknowledgements}
This work is supported by  the
Grant-in-Aid for  Scientific Research Fund of the Ministry of 
Education, Science and Culture of Japan No.22540274, the Grant-in-Aid
for Scientific Research (A) (No. 22244030), the
Grant-in-Aid for  Scientific Research on Innovative Area No.21111006,
and the Grant-in-Aid for the Global COE Program 
``The Next Generation of Physics, Spun from Universality and Emergence". 

%

\end{document}